\documentclass[12pt]{article}
\begin{document}
\title{Doppler effect and frequency-shift in optics}
\author{S.C. Tiwari \\
Institute of Natural Philosophy \\
1 Kusum Kutir, Mahamanapuri \\
Varanasi 221005, India }
\maketitle
\begin{center}
\textbf{{\large{}Abstract}}
\end{center}

A critical review of frequency-shift phenomena a la Doppler effect is presented. 
The importance of Fermi's theory of 1932 is pointed out,and it is argued that there 
exist gaps in our understanding of this phenomena at a fundamental level. Alternative 
mechanism in terms of photon number oscillations is suggested for polarization 
changing experiments. The physical reality of single photon is revisited with a new
interpretation of zero point energy. Total energy of the photon comprises of translational
(wave-like) and rotational (extended particle-like) energy.

\section{Introduction}

Doppler effect is a well established phenomenon, and has found many applications. Physical mechanism of 
Doppler effect for sound waves is very transparent as explained in elementary textbooks, however for 
light waves after the advent of the theory of relativity and rejection of aether hypothesis there arise subtle problems
 in its interpretation. The most cogent argument for Doppler effect in the case of electromagnetic (EM) 
waves is based on the invariance of the phase of the wave (treating frequency and wave vector as a 
four-vector) under Lorentz transformation. Considering the fact that all the known mechanisms to change 
the frequency of a monochromatic EM wave depend on a dynamic interaction process or an active 
circuit(or optical) element, purely kinematic origin of frequency shift is intriguing. If the claim is 
made for a single photon, the physical explanation becomes obscure; in fact the word photon is used
 rather casually in the literature, see a critique in \cite{1}. In the present paper, we analyze fundamental 
issues involved in the phenomena of frequency shifts for light waves a la Doppler effect and the 
generalizations.

In the next section, standard Doppler effect is defined, and frequency-shift phenomena are introduced. 
A critique on the relativistic treatment and the quantum theory of Doppler effect is presented in Sec. 3. 
Basic issues on phase and amplitude are discussed in Sec.4, and an alternative interpretation of 
frequency shift is suggested. In Sec.5 a picture of single photon is envisaged in which translatory
periodic motion and internal rotation in the tranverse plane represent the photon.
It is argued that total energy of photon is equally divided between linear motion and spin. 
An outline of experimental test for the alternative interpretation at single photon level 
is presented. Concluding remarks constitute the last section.

\section{Doppler effect and the generalizations }

Relative motion between the source generating a wave and the observer receiving the wave leads to a 
change in the frequency of the wave-such a kinematic effect is known as Doppler effect. Obviously to 
define relative motion one has to introduce inertial frame of reference (IFR). The weakness of 
gravitational interaction allows one to assume that IFR could be reasonably defined. Since sound 
wave is a mechanical disturbance that propagates in a material medium, IFRs for source, observer 
and wave are easily defined, and a moving source (or an observer) counting different frequency 
makes physical sense. In the case of light, the postulated EM disturbance propagates in vacuum 
with constant wave velocity, c i.e. the velocity c is independent of the IFRs. Light emitted from 
moving (stationary) source is shifted in frequency as observed by a stationary (moving) observer 
given by the relation
\begin{equation}
\nu_0=\gamma\nu_e(1-\beta\cos{\theta})						
\end{equation}
\begin{equation}
\tan\theta^\prime=\frac{\sin\theta}{\gamma(\cos\theta-\beta)}
\end{equation}
Here $\nu_{\mathrm{o}}$ and $\nu_{\mathrm{e}}$ are the observed and emitted frequencies of light 
respectively,$\beta$ = v/c,$\gamma =\sqrt(1-\beta^2)$, \textbf{v} is the relative velocity between 
the source and the observer, and the wave vectors of light (in different IFRs) make angles $\theta$ 
and $\theta^\prime$ with respect to \textbf{v}. To first order in $\beta$ with $\theta$ =0,  the 
standard Doppler shift is obtained from Eq. (1)     
\begin{equation}
\nu_0=\nu_e(1-v/c)										
\end{equation}
There is a frequency shift in the direction perpendicular to the motion of the source, termed as 
transverse Doppler effect. For $\theta  = \pi$ /2 , Eq. (1) gives
\begin{equation}
\nu_0=\nu_e\gamma
\end{equation}

Note that both linear and transverse Doppler effects arise due to the translational motion of the bodies, 
let us designate them by TDE (translational Doppler effect). Garetz \cite{2}  introduced the term angular 
Doppler effect (ADE) to denote frequency shift caused by the rotation of the light emitting body such 
that the axis of rotation and the direction of propagation of light are parallel. The treatment given
 by Garetz is applicable to polarized plane waves, while Allen et al \cite{3} consider finite-sized 
orbital angular momentum (OAM) carrying beams to obtain azimuthal Doppler shift analogous to ADE. The 
notion of rotational frequency shift (RFS) was discussed in 1997 \cite{4} based on non-relativistic 
quantum mechanical treatment for a rotating radiating two-level system interacting with quantized 
EM field. Authors of \cite{4} note that, ``The RFS should not be confused with the ordinary linear 
Doppler shift observed for rotating objects (for example, stars or galaxies) that is due to the 
instantaneous linear motion of the emitter. This linear Doppler shift is maximal in the plane of 
the rotation while the RFS is maximal along the angular velocity, that is in the direction perpendicular 
to the instantaneous velocity. Thus the RFS competes with the quadratic Doppler shift rather than 
with the linear Doppler shift''. Note that the frequency shifts caused by moving reflecting mirrors 
or rotating wave plates have the origin in the Doppler effect (TDE or ADE). Garetz \cite{2} gives 
a nice discussion based on Beth experiment for interpreting frequency shift caused by rotating 
half-wave plates \cite{5}.

In his letter, Garetz \cite{2} also interprets rotational Raman spectra and fluorescence doublets in 
terms of ADE. There is another class of frequency-shift phenomenon that occurs in the non-inertial 
frames. It is well known that the gravitational force in general theory of relativity is postulated 
to be an effect of curved pseudo-Riemannian space-time, therefore one can anticipate gravitational 
red shifts. In a simple example \cite{6}, one can construct a space-time metric for a rotating 
coordinate system with a uniform angular velocity of rotation. Assuming synchronization of clocks 
at each instant of time the shift in frequency of light can be calculated. In general, the difference 
in the gravitational potential at two different points determines the gravitational frequency shift. 
In 1960, Pound and Rebka verified gravitational redshift for gamma rays using Mossbauer effect, 
and Hay et al measured the frequency shift of 14 KeV gamma rays in a rotating system \cite{7}.

\section{Frequency shifts-physical mechanisms}
\subsection{Simple explanations}

The standard approach to Doppler effect is based on the invariance of the phase factor 
(\textbf{k.r} -$\omega$ t)\textbf{ }under Lorentz transformation from one IFR to another. 
Jackson observes \cite{8} that, `the phase of a wave is an invariant quantity because the phase 
can be identified with the mare counting of wave crests in a wave train, an operation that must 
be the same in all inertial frames'. In \cite{8} the second postulate of Einstein's special
 relativity is simply stated as `speed of light is independent of its source'. The question 
arises: what  is the meaning of observing `a wave train of light' in a specific IFR? Dingle 
in a short monograph \cite{9} touches upon this problem to some extent. He says that the 
coordinates appearing in the phase of the wave train are in a frame in which the source of 
light is at rest, and notes that, '' It is not necessary, of course, that we should think 
of this equation as representing a wave motion in the space between the source of light and 
the observer, although it is frequently convenient to do so. The facts of observation are all 
equally well expressed if we regard the various quantities as representing something 
characterizing the source of light itself ''. His discussion on the question whether the frequency 
change is an intrinsic property of source or not is interesting, however the physical mechanism 
for the Doppler effect is not clarified. Note that the speed c represents the motion of the 
constant phase surfaces independent of IFRs. The literature does not address the problem: to 
which physical system, i.e. the source or the plane EM wave, the  coordinates (\textbf{r}, t) 
and the quantities  (\textbf{k},$\omega$ ) belong. Merely counting of wave crests is not enough.

Einstein in his 1905 paper \cite{10} claims that aether is superfluous since an 'absolutely
stationary space' provided with special properties is not required, and it is not necessary
to introduce a velocity-vector to a point in the empty space in which the electromagnetic
processes occur. Note that he does not give physical meaning to frames of reference in relative
uniform translatory motion, and the definition of the velocity of light, namely, light path
divided by time interval is too simplistic. The second postulate of relativity says, 'Any ray
of light moves in a stationary system of coordinates with the determined velocity c, whether the
ray be emitted by a stationary or a moving body'. In Sec.7 of his paper a theory of Doppler
effect is given. Following Einstein let us consider a frame K in which a source of light emits
waves, and an observer is moving with velocity v relative to infinitely distant source of light.
Next the connecting line source-observer is considered where an observer is at rest in the 
moving system K. The space-time coordinates in the phase factor are assumed to be that of the
source frame, and while the wave reaches the observer traversing in vacuum with constant
velocity the space-time coordinates of the phase factor are now changed to that of the
observer frame. Invariance of phase immediately leads to the Doppler shift and aberration.
The most puzzling aspect is that intervening inert empty space between source and observer
somehow is capable to affect the internal constitution of light i.e. apparently no
physical mechanism is needed to change the frequency of the light wave. Relativistic
world view merely asserts that space and time are not absolute and the phase invariance
from one frame to another changes the space-time coordinates and correspondingly changes
in frequency and wave vector.

Alternative to wave theory, particle picture of light is used in \cite{2}. Though Garetz cites 
Sommerfeld's book \cite{11}, Doppler effect from `photon point of view' seems to have been given 
by Schrodinger \cite{12}. Fermi's 1932 article \cite{13} presents this approach nicely, and gives 
a quantum theory of Doppler effect using \textbf{p.A} form of atom-radiation interaction. 

Following Fermi, let us consider a two-level quantum system, e.g. an atom with energy levels
E$_{\mathrm{1}}$ and E$_{\mathrm{2}}$. Let
\begin{equation}
h\nu   =  E_2-E_1									
\end{equation}
Assuming the atom to be at rest, the emitted frequency is $\nu_{\mathrm{e}}$. Supposing the 
atom to be in the excited state, E$_{\mathrm{2}}$, and moving with velocity v, the frequency 
of the emitted radiation is calculated using the energy and momentum conservation. A little algebra 
finally leads to the Doppler shift, Eq. (3). It appears that there is an inconsistency. Why should 
there be no recoil for the atom in the rest frame? The conservation of momentum gives
\begin{equation}
mu = - h\nu^\prime /c									
\end{equation}
where $u$ is the velocity acquired by the atom initially at rest. The initial energy of the atom is 
E$_{\mathrm{2}}$ (being in the excited state), and after emission of radiation it is 
E$_{\mathrm{1}}$ + (1/2)mu$^{\mathrm{2}}$. Thus the emitted energy is 
\begin{equation}
h\nu^\prime  =  E_2-E_1-\frac{1}{2}mu^2							
\end{equation}
Evidently $\nu^\prime$ is not equal to $\nu_{\mathrm{e}}$, and there is a shift in frequency. It is 
only when recoilless emission (i.e. Mossbauer like effect) is enforced that the emitted frequency is 
exactly $\nu_{\mathrm{e}}$.

\subsection{Quantum theory}

Dirac's quantum theory of radiation finds beautiful application to explain classical phenomena like light 
propagation in vacuum and the Lippman fringes in Fermi's paper \cite{13}. Here we present the main 
ingredients of this theory as applied to the Doppler effect. The atom, the radiation field and the
 atom-radiation interaction are treated as a single system. The radiation field enclosed in a finite
 volume of space, V is represented by quantized oscillators in the plane wave approximation. The
 interaction term is obtained by the transformation \textbf{p}  $\to$ 
\textbf{p}-e \textbf{A}, and assumed to act as a small perturbation. In the nonrelativistic 
formulation, the standard perturbation theory is used to calculate the time evolution of the 
probability amplitudes of the Schrodinger wave function. In the Doppler shift phenomenon, a simple 
hydrogen-like atom is considered. The Hamiltonian for this system is given by
\begin{equation}
H=\frac{p_1^2}{2m_1}+\frac{p_2^2}{2m_2} + U(\textbf{q}_1 - \textbf{q}_2)+ \sum_s
\frac{1}{2}(p_s^2 + \omega_s^2 q_s^2) - \frac{e}{m_1}\textbf{A}(\textbf{q}_1).
\textbf{p}_1+ \frac{e}{m_2} \textbf{A}(\textbf{q}_2).\textbf{p}_2
\end{equation}
The interaction term proportional to A$^{\mathrm{2}}$ is neglected. 
Here \textbf{q}$_{\mathrm{1}}$(\textbf{q}$_{\mathrm{2}}$), m$_{\mathrm{1}}$(m$_{\mathrm{2}}$) 
and \textbf{p}\textbf{$_{\mathrm{1}}$}(\textbf{p}\textbf{$_{\mathrm{2}}$}) are respectively 
the coordinate, mass and momentum of proton (electron); U is the Coulomb binding energy. The 
fourth term in H represents quantized radiation, H$_{\mathrm{r}}$, and the last two terms 
represent the interaction Hamiltonian, $H^\prime$. The vector potential is given by
\begin{equation}
\textbf{A}=c\sqrt{\frac{8\pi}{V}}\sum_s ({\hat e}_s q_s \sin\Gamma_s)
\end{equation}
\begin{equation}
\Gamma_s =\textbf{k}_s.\textbf{r}+\beta_s
\end{equation}
Here \textbf{e}$_{\mathrm{s}}$ is a unit vector along \textbf{A}, \textbf{k}\textbf{$_{\mathrm{s}}$} 
is the propagation vector, and $\beta_{\mathrm{s}}$ is a constant phase.

Fermi makes two points: in order to account for the impulse of the radiation, the center of mass 
motion has to be incorporated in the theory, and the usual assumption of the constancy of the phases. 
$\Gamma_{\mathrm{s}}$ does not hold in this case. Defining new variables, namely the center of 
mass coordinate \textbf{R} and relative coordinate \textbf{q}, we have
\begin{equation}
\textbf{R}= \frac{m_1\textbf{q}_1 +m_2\textbf{q}_2}{M}
\end{equation} 
\begin{equation}
\textbf{q}=\textbf{q}_1 -\textbf{q}_2
\end{equation}
Here the total mass is M = m$_{\mathrm{1}}$ + m$_{\mathrm{2}}$. The corresponding momenta are given by
\begin{equation}
\textbf{P}=\textbf{p}_1 +\textbf{p}_2								
\end{equation}
\begin{equation}
\textbf{p}= \frac{m_2\textbf{p}_1-m_1\textbf{p}_2}{M}
\end{equation}
The Hamiltonian of the system in new variables is transformed to the unperturbed part
\begin{equation}
H_0=\frac{P^2}{2M}+ \frac{p^2}{2m}+U(\textbf{q})+H_r
\end{equation}
and the interaction part
\begin{equation}
H' = -\frac{e}{m}\sqrt{\frac{8\pi}{V}}\sum_s \bf{{\hat e}_s}.\textbf{p} q_s sin\Gamma_s	
\end{equation}
Note that m = m$_{\mathrm{1}}$m$_{\mathrm{2}}$/M, and the assumption is made that the phases 
$\Gamma_{\mathrm{s}}$ depend only on \textbf{R}, i.e. 

\begin{equation}
\Gamma_s=\textbf{k}_s. \textbf{R}+\beta_s					
\end{equation}
In Eq. (15), the first term represents the gross motion of the atom, and the next two terms define 
internal states. Assuming that the wave function of the system can be written as a product of the
 eigenfunctions of gross motion, internal state and the radiation we have
\begin{equation}
\Psi=\Psi_{cm}\Psi_{int}\Psi_r									
\end{equation}
and the energy eigenvalue is given by the sum of the corresponding eigenvalues. The probability 
amplitudes will satisfy a first order (time) differential equation, and involve the matrix elements
 for the interaction H$\prime$. The matrix elements nicely separate into a product of integrals 
with respect to \textbf{R}, \textbf{q }and\textbf{ }q$_{\mathrm{s}}$. The integral involving 
\textbf{R} looks like
\begin{equation}
I =  \int{e^{-i\textbf{K}_n.\textbf{R}}sin(\textbf{k}_s.\textbf{R}+\beta_s)e^{i\textbf{K}_m.
\textbf{R}}d\textbf{R}}											
\end{equation}
Here $\hbar$\textbf{K} is the momentum of the atom. Exponential representation of sine function 
immediately leads to the momentum conservation law
\begin{equation}
{\bf K}_m - \textbf{K}_n \pm \textbf{k}_s=0							
\end{equation}
Together with the energy conservation, this equation leads to the Doppler shift as explained in the
 preceding sub-section.

\subsection{Critique}

During past more than a decade, there has been a great deal of activity on the gross motion of atoms 
subjected to laser light beams. The theory of a moving two-level atom interacting with the light 
beam adequately brings out the physics of the phenomena such as atomic beam deflection and laser 
cooling. In the cold atom optics, at very low temperatures where the atomic de Broglie wavelength 
is of the same order as the wavelength of the light, the gross atomic motion needs to be quantized. 
We refer to \cite{14} for basic theory and original references cited therein. Allen et al \cite{3}
generalize the canonical theory for the radiation-pressure effects on the gross motion of the 
atom \cite{15} replacing the plane wave representation by the Laguerre-Gaussian modes to obtain 
the azimuthal Doppler shift. Let us consider the Hamiltonian given in \cite{3} for a two-level
 atom of resonant frequency $\omega_{\mathrm{o}}$, interacting with light mode of frequency 
$\omega$
\begin{equation}
H=\hbar\omega\pi^\dagger\pi+\frac{P^2}{2M}+U(\textbf{R})+\hbar\omega a^\dagger a-i\hbar[\pi^\dagger 
a f(\textbf{R})-f^*(\textbf{R})a^\dagger\pi]								
\end{equation}

Here $\pi$ and $\pi^{\dagger}$ are ladder operators for internal two-levels, and a and a$^{\dagger}$ 
are the annihilation and creation operators respectively of the radiation field. The operator  
f(\textbf{R} ) arises from the dipole interaction term -e\textbf{E.r}. Comparing Eq. (21) with 
Fermi's theory, Eqs. (15) and (16) it is straightforward to recognize that for a single mode 
radiation field both have nice correspondence except that Fermi uses \textbf{p.A} interaction term. 
Note that the operators q$_{\mathrm{s}}$ and p$_{\mathrm{s}}$ in the radiation field Hamiltonian 
can be expressed in terms of non-Hermitian operators a and a$^{\dagger}$
\begin{equation}
q_s=\sqrt{\frac{\hbar}{2\omega}}[a_s+a^{\dagger}_s]
\end{equation}
\begin{equation}
p_s=i\sqrt{\frac{\hbar\omega}{2}}[a^{\dagger}_s-a_s]
\end{equation}
Thus it would seem that the quantum theory of Doppler effect has been rediscovered. Regarding 
the choice of dipole interaction term, see \cite{16} for a critical analysis of the problem.

Quantum theory of rotational effects given in \cite{3} seems quite logical. On the other hand, the 
theory given in \cite{4} is based on an electron bound by a time dependent potential V(\textbf{r}(t)) 
where the coordinate \textbf{r} does not bring out the gross motion clearly. At a fundamental level, 
it is true that the separation into internal and external parts of a system is unsatisfactory, but
 the claim that a uniform translation does not affect internal energy level difference while a 
uniform rotation does, remains rather obscure in \cite{4}. In fact, it is not clear if RFS is 
equivalent to the azimuthal Doppler shift. In a recent report \cite{17} it has been claimed
that the RFS has been observed for the first time in a molecular system.

Let us contrast the relativistic explanation and quantum theory of the Doppler effect. The Lorentz 
transformations and the invariance of phase (\textbf{k.r} - $\omega$t) under them do not throw any 
light on the mechanism of energy-momentum transfer that clearly arises in the quantum theory. 
If we consider the phase factor $\Gamma_{\mathrm{s}}$ in quantum theory then it would appear that 
the coordinates in (\textbf{k.r} - $\omega$t) should correspond to the source. It is intriguing 
that in the relativistic explanation, one merely asserts that let there be an IFR moving with a 
velocity \textbf{v} relative to another IFR, but does not point out the physical process by which a 
stationary object acquires uniform motion. Similar to this, in quantum theory, one asserts the 
collapse of the wave function to a certain eigenstate. In this sense, both relativistic and quantum 
theories are unsatisfactory. It is interesting to note that Einstein in his 1917 paper on
quantum theory of radiation \cite{18} justifies the necessity of classical EM theory saying:
'Whatever the form of the theory of electromagnetic processes, surely in any case the Doppler
principle and the aberration will remain valid'. In the background of the developments preceding
relativity paper and the formative years of quantum theory such an obscure and paradoxical
approach to Doppler effect seems understandable, however with our present knowledge of
light, photon and light-matter interaction we must address the fundamental issues afresh.

A careful examination of quantum theory shows that the transition of the atomic internal state is
 not instantaneous, there is a finite lifetime. In relativity, the time dilation is often used 
to explain the longer lifetime of a moving unstable particle as compared to the static one. In our 
previous work \cite{19} we argued that kinematic explanation for an irreversible process posed a 
paradox. Critical analysis of the concept of time in relativity led us to suggest that the 
different IFRs are characterized by differing constant potentials. Similar to the Aharonov-Bohm 
effect or geometric phase, under suitable conditions the observable effects of constant potentials 
become manifest: change in lifetime and Doppler effect could be such phenomena \cite{20}. 
We refer to \cite{21} for detailed exposition of the ideas on time.

Preceding discussion shows that the quantum theory of Doppler shift takes into account the effect of radiation 
pressure on the gross quantized motion of the atom. It is also known that the lifetime of a two-level 
system can be calculated using Fermi's golden rule \cite{13}. We suggest that a self-consistent
 theory treating both the Doppler effect and lifetime dilation simultaneously would be a great
 advancement. A possible approach being investigated is based on the incorporation of Berry's 
connection and non-equivalence of IFRs for a two-level atom plus radiation system. Let us recall
 that the standard theory for a quantum system is often simplified in terms of two sets of variables.
 First, one set of variables is kept constant, and the problem is solved for the other set of 
dynamical variables. Next the first set of variables is varied to obtain the effective dynamics
 \cite{22}. The induced vector potential or Berry's connection is shown to arise 
 in the effective dynamics. Wilkens in an interesting paper \cite{23} considers spontaneous
  emission of a moving atom including the Roentgen interaction term in the Hamiltonian. Here the
  magnetic dipole-like term arises due to the motion of the radiating atom. Cresser and
  Barnett \cite{24} elucidate the basic problem considering a moving electric dipole that
  generates a magnetic dipole moment and the Doppler shift, and argue that these physical
  processes 'conspire' to give time dilation effect for the lifetime of the decaying atom
  as demanded by special relativity. Is it not strange that time dilation kinematic effect
  is unreservedly accepted for a decaying irreversible process? Our suggestion that time
  dilation effect is an analogue of geometric phase effect becomes plausible in the
  light of Wilken's work \cite{25}. Similar to the Aharonov-Bohm effect one expects
  topological phase for a moving dipole \cite{25} ,see also \cite{26}. Another related interesting
  result in quantum optics is that of the enhancement or inhibition of spontaneous
  emission if the surrounding quantum vacuum is modified \cite{27}. In a somewhat different 
  context it has been argued \cite{28} that Doppler shift and broadening of spectral lines
  both could be explained based on Wolf effect.

\section{Phase and amplitude of light wave}

In the review on angular momentum of light \cite{29}, a brief discussion is given on the RFS. A 
rotating half-wave plate shifts the frequency of a circularly polarized light wave, while for a 
circularly polarized Laguerre-Gaussian modes a rotating Dove prism and half-wave plate give rise 
to the frequency shift. A recent review \cite{30} dwells upon the energy exchange mechanism for the 
frequency shifts, while the frequency shift in the context of geometric phase was suggested by 
Simon et al \cite{31}. In fact, the significance of energy conservation for Doppler shift was 
clear in Fermi's analysis \cite{13}, and for the angular Doppler effect, Garetz \cite{2} considers 
the torque for a rotating wave plate, and the work done on or by the wave plate for the changes 
in the spin angular momentum of light. Note that the experiments discussed in \cite{2} are 
carried out for the classical light beams though the author uses photon point of view. In the 
preceding section we have analyzed the physical mechanisms responsible for the Doppler 
frequency shifts, and pointed out that atom-radiation interaction is crucial for frequency 
changes of radiation though some questions remain unsettled. Could the frequency of a monochromatic 
light wave be changed by rotating wave plates? If the frequency of a photon is its intrinsic 
attribute, could it be changed by a rotating optical element? We confine our attention to the 
polarization changing effects in the following, and seek an alternative interpretation of the 
experimental results.

The experiments show that the intensity of light transmitted through a rotating wave plate 
behaves sinusoidally with time. For an ideal case of lossless transmission and perfect fringe 
visibility, we have
\begin{equation}
I_t =I_i[1+cos(\theta_0+2\Omega t)]									
\end{equation}

The wave plate is assumed to rotate at an angular frequency of $\Omega$, and the time-dependence 
of output intensity, I$_{\mathrm{t}}$ is interpreted as a frequency shift of $\pm$ 
2$\Omega$. We argue that a time-dependent phase does not necessarily imply the frequency change. 
For a plane monochromatic EM wave, the electric field vector can be written as
\begin{equation}
\textbf{E}=\textbf{E}_0 e^{i(\textbf{k.r}-\omega t)}							
\end{equation}

The Poynting vector, \textbf{S}, is in the direction of \textbf{k}, normal to both electric and 
magnetic field vectors. The magnitude of \textbf{S} averaged over time period gives the intensity 
of the plane wave.

\begin{equation}
I=\frac{c}{8\pi}\textbf{E}_0.\textbf{E}_0^*								
\end{equation}
It is clear that the intensity does not depend on the frequency of the wave, however starting 
with the vector potential \textbf{A} and applying the transversality condition (i.e. radiation gauge)

\begin{equation}
\textbf{A}=\textbf{A}_0 e^{i(\textbf{k.r}-\omega t)}						
\end{equation}
The expression for the intensity in terms of \textbf{A}\textbf{$_{\mathrm{o}}$} is easily calculated to be 
\begin{equation}
I=\frac{\omega^2}{2\pi c}\textbf{A}_0.\textbf{A}_0^*				
\end{equation}
Eq. (28) shows the frequency dependence of the intensity. Recall that the total energy i.e. the sum of 
kinetic and potential energy of a classical simple harmonic oscillator is given by 
\begin{equation}
E_{tot}^c=\frac{1}{2} m\omega^2 x_m^2	
\end{equation}
where x$_{\mathrm{m}}$ is the amplitude, and  $\omega$ is the angular frequency of the oscillator. 
Note the similarity of the expression of the intensity I vide Eq. (28) with E$_{\mathrm{tot}}$.

Returning to the problem of rotating wave plate, light wave propagating along, let us say z-axis, 
can be represented by a row vector [E$_{\mathrm{x}}$   E$_{\mathrm{y}}$)]and its Hermitian adjoint 
by a column vector. The effect of an optical element using Jones calculus can be described in terms 
of a 2x2 transmission matrix. A rotating half-wave plate introduces a time dependent phase shift 
depending on $\Omega$. Bretanaker and LeFloch \cite{32} rightly note that Jones calculus leads 
to the observed phase shift, and seek angular momentum exchange mechanism to interpret the frequency 
shift. Authors invoke energy exchange using the energy of N photons equal to N$\hbar\omega$, 
and change in  $\omega$ due to rotating wave plate. In laser physics, one uses the notion of 
energy flux in terms of the number of photons, N
\begin{equation}
U_f=\frac{cN\hbar\omega}{V}							
\end{equation}
Assuming that the total number of photons in the light wave consists of spin + $\hbar$ , 
N$_{\mathrm{+}}$ and spin $-\hbar$ , N$_{\mathrm{-}}$, we suggest that the time-dependent phase 
arises due to the modulation of photon numbers in contrast to the frequency change 
suggested in \cite{32}.

	Since the experiments measure the intensity oscillations, it is not obvious how to 
distinguish the photon number oscillations from the frequency shifts. Novel experimental scheme
 would have to be devised that markedly depends on the number of photons in the light wave. 
The experiments based on photo-electric effect is one of the possibilities however quantum theory 
of radiation shows that the photon number operator is not directly observable \cite{13} p.629, 
therefore it is not clear whether it would lead to conclusive results.

\section{Single photon}
Instead of classical light wave, the frequency-shift phenomena at a single photon level appears to
 be an attractive idea. The first fundamental issue in this case is, of course, whether photon has 
physical reality or it is merely a mathematical construct in the form of vacuum excitation. Lamb has 
severely criticized the concept of photon used by `the laser community' \cite{33}, however for a
 balanced critique we refer to \cite{1}. In quantum optics, it is asserted that the radiation enclosed in
 a cavity has discrete electric field with an amplitude of 
\begin{equation}
E_{0r}=\sqrt{\frac{\hbar\omega}{2V}}								
\end{equation}

and for n photons the quantum of electric field is E$_{\mathrm{or}}$ $\sqrt{n}$. In \cite{34} Knight 
draws attention to the experiments that seem to test the consequences of the quantized electric field.
 In spite of the recent advances in quantum optics, some fundamental questions have remained 
unresolved; besides the problems reviewed in \cite{1} we refer to a passionate critique on the 
Copenhagen interpretation of quantum mechanics by Post \cite{35}. Of particular significance is 
the anticipation of a zero-point energy by Planck in 1912 discussed in detail in \cite{35}. 
Randomness in the mutual phases of an ensemble of classical harmonic oscillators was considered
 by Planck, and for thermodynamic equilibrium an average zero-point energy of $\hbar\omega/2 $ 
per oscillator was necessary. To summarize: the meaning of electric field amplitude and phase 
for a single photon deserve a careful attention whether one rejects or accepts the physical 
reality of photon \cite{36}.

	To gain insight, we revisit elementary considerations. Note that the electric field 
amplitude E$_{\mathrm{or}}$ depends on the frequency of the radiation. Eqs. (26) and (30) 
do indicate such a relationship, but there maybe a deeper reason at the quantization level. 
The canonical variables (q$_{\mathrm{s}}$, p$_{\mathrm{s}}$) are expressed in terms of creation 
and annihilation operators vide Eqs. (22) and (23), and the frequency dependent factors are 
absorbed in the definitions. Essentially similar step appears in quantum oscillator that has 
total energy
\begin{equation}
E^q_{tot} =(n+1/2)\hbar\omega										
\end{equation}

In the textbooks, it is usually proved that for any value of n, the  expectation values of kinetic 
and potential energy are equal and each one is half of the total energy just as in the case of
 classical oscillator. However, the crucial difference between the classical and quantum 
oscillators is not mentioned, namely the absence of an amplitude factor in Eq. (32). The 
expression (31) for E$_{\mathrm{or}}$ is based on the calculation of the expectation value 
of the square of the field operator since 
\begin{equation}
<n|{\hat E}|n> = 0
\end{equation}

In the case of a classical oscillator the amplitude of x is x$_{\mathrm{m}}$, while the 
momentum amplitude is m $\omega$ x$_{\mathrm{m}}$. Assuming that the product of the two is 
constant, and setting it to be $\hbar$ we get 

\begin{equation}
m\omega x^2_m =\hbar
\end{equation}
The energy of the oscillator becomes
\begin{equation}
E^c_{tot} =\hbar\omega/2
\end{equation}

In this form, the zero-point energy corresponds to a single oscillator unlike the randomized 
phases for the ensemble of oscillators in Planck's work or vacuum quantum fluctuations in 
quantum optics. Though as yet no physically sound and concrete model of photon has been 
developed, it is generally believed that photon has energy $\hbar\omega$, momentum h$\nu$/c, 
and spin angular momentum $ \pm \; \hbar$. I have never been able to understand 
why a simple natural question is not asked:What is the energy of the photon associated
with its spin angular momentum? It is quite possible that it may have something to do
with Einstein's reluctance to attach much significance to photon spin \cite{36}
and for historical reason that quantum optics formalism takes care of spin as 
polarization index for all practical purposes. In the light of expression (35), let us visualize 
photon as an object possessing internal rotational motion of an extended structure, 
and translational motion as a periodic propagating disturbance. Recall that the kinetic 
energy for a classical particle having momentum $p$ is $(1/2) pv$, and the rotational energy 
is equal to $ (1/2) L\omega$, where $L$ is the angular momentum. Now we split the 
energy  $\hbar\omega$ of photon as follows:
\begin{equation}
\hbar\omega= \frac{1}{2}\hbar\omega+\frac{1}{2}(\frac{h\nu}{c})c	
\end{equation}

Note that the expression (36) looks like $(1/2) L\omega + (1/2) pv$. Further the translational 
periodic motion as a harmonic oscillator would have the energy $(1/2) \hbar\omega$, Eq. (35) 
that would be consistent with the second term on the right hand side of Eq. (36). 
In \cite{36} it has been speculated that internal and external periodicities are in 
synchronization for the extended structure of photon moving in vacuum.

An important significance of the present photon model is on the wave-particle duality. 
It is known that the first enunciation of the complementarity principle by Bohr was based on the
simple argument:energy and momentum represent particle attributes while frequency and 
wavelength represent extended wave, therefore, the Planck-Einstein relationship
between them imply duality. In the present model the extended structure transverse to
the direction of the propagation has particle-like attribute, and the periodic translatory
motion would give rise to wave-like effects. The radical revision of the classical picture
envisaged here is that of discarding the description based on point particle and 
instantaneous dynamical variables. Just as zero point energy is not some mysterious quantum
vacuum effect, the Heisenberg's uncertainty relation merely represents finite discrete
spatial and temporal units such that the product of the size of the photon and its momentum
is equal to the Planck constant.

The intriguing questions regarding the frequency dependent amplitude, photon number 
oscillations, and frequency shifts will assume distinct significance for single photon 
experiments i.e. the transmission of a photon through a rotating wave plate. The time 
scales in the interference experiments would be most crucial. In the context of
our photon model the polarization-dependent experiments would show particle behavior,
while the momentum exchange experiments would show wave-like aspects. We expect
important implications on tha black body radiation physics which will be discussed
elsewhere.

\section{Conclusion}

In this paper, we have critically reviewed the physical mechanism for the Doppler shift, 
and drawn attention to Fermi's quantum theory that considered quantized gross motion of 
the atom more than seven decades ago. In spite of the well known relativistic explanation 
and quantum theory of Doppler effect, we point out that there exist gaps in our 
understanding of this phenomenon. We suggest a self consistent theory taking into 
account the atom-radiation interaction for life-time dilation of a moving atom and 
the Doppler shift will provide new insights.

In the case of frequency-shift phenomena observed using passive optical elements 
(e.g. rotating wave plates), we offer an alternative mechanism in terms of the photon 
number oscillations for the polarization changing experiments. Some intriguing aspects 
of amplitude and phase of the light wave are discussed, and implications on the 
physical model of photon are indicated. A conclusive proof whether the intensity 
oscillations observed experimentally are due to frequency-shift or time-dependent 
amplitude due to photon number oscillations appears difficult, however at a single 
photon level it should be possible to distinguish the two.

Different strands on the frequency shift phenomena for electromagnetic radiation
indicate the need for replacing the obscure and counter-intuitive physics
of the Doppler effect by an underlying simple unifying principle. I believe this
principle will emanate from the understanding of the physics of aether or quantum vacuum
or what I prefer to call manifest dynamical space \cite{37}.

\section{Acknowledgements}

I am grateful to Prof. L. Allen for encouragement and stimulating correspondence on 
frequency-shift phenomena. I thank Professors Sisir Roy, Peeter Saari and A. Luis for their 
interest in my work. The Library facility of the Banaras Hindu University, Varanasi 
is acknowledged.

\end{document}